\documentstyle[aps,prl,epsf,amsfonts]{revtex}
\newcommand{\eqb}{\begin{equation}}
\newcommand{\eqe}{\end{equation}}
\newcommand{\dmb}{\begin{displaymath}}
\newcommand{\dme}{\end{displaymath}}

\newcommand{\ep}{\varepsilon}
\newcommand{\eab}{\begin{eqnarray}}
\newcommand{\eae}{\end{eqnarray}}
\newcommand{\ra}{\right\rangle}
\newcommand{\la}{\left\langle}
\newcommand{\e}{\mbox{e}}

\newcommand{\cD}{{\cal D}}
\newcommand{\nc}{\newcommand}
\nc{\vivi}{very interesting and very important}
\nc{\al}{\alpha}
\nc{\ga}{\gamma}
\nc{\de}{\delta}
\nc{\ze}{\zeta}
\nc{\et}{\eta}

\nc{\Th}{\Theta}
\nc{\ka}{\kappa}
\nc{\lam}{\lambda}
\nc{\rh}{\rho}
\nc{\si}{\sigma}
\nc{\ta}{\tau}
\nc{\up}{\upsilon}
\nc{\ph}{\phi}
\nc{\ch}{\chi}
\nc{\ps}{\psi}
\nc{\om}{\omega}
\nc{\Ga}{\Gamma}
\nc{\De}{\Delta}
\nc{\La}{\Lambda}
\nc{\Si}{\Sigma}
\nc{\Up}{\Upsilon}
\nc{\Ph}{\Phi}
\nc{\Ps}{\Psi}
\nc{\Om}{\Omega}
\nc{\ptl}{\partial}
\nc{\del}{\nabla}
\nc{\be}{\begin{equation}}
\nc{\ee}{\end{equation}}
\nc{\bea}{\begin{eqnarray}}
\nc{\eea}{\end{eqnarray}}
\nc{\ov}{\overline}
\nc{\gsl}{\!\not}

\newcommand{\bi}[1]{\bibitem{#1}}
\newcommand{\fr}[2]{\frac{#1}{#2}}
\newcommand{\gm}{\mbox{$\gamma_{\mu}$}}

\newcommand{\gf}{\mbox{$\gamma_{5}$}}

\newcommand{\qq}{\langle \ov{q}q\rangle}
\draft

\begin{document}

\title{$\rho$-$\omega$ mixing in 
asymmetric nuclear matter via QCD sum rule approach}
\author{Abhee K. Dutt-Mazumder}
\address{Physics Department, McGill University, 3600 University St,
Montreal,Quebec H3A 2T8, Canada}
\author{Ralf Hofmann and Maxim Pospelov}
\address{Theoretical Physics Institute,  
         University of Minnesota,  
         Minneapolis, MN 55455, USA }

\maketitle

\begin{abstract}

 
We evaluate the operator product expansion (OPE) for a 
mixed correlator of the isovector and isoscalar vector currents
in the background of the nucleon density with intrinsic isospin asymmetry 
[i.e. excess of neutrons over protons] and match it with its imaginary part, 
given by resonances and continuum, via the dispersion relation. 
The leading density-dependent contribution to $\rho-\omega$ mixing 
is due the scattering term, which turns out to be larger than 
any density dependent piece in the OPE. We estimate that the 
asymmetric density of $n_n-n_p \sim 2.5 \times 10^{-2} ~{\rm fm^3}$
induces the amplitude of $\rho-\omega$ mixing, equal in 
magnitude to the mixing amplitude in vacuum, 
with the constructive interference for positive and destructive for negative 
values of $n_n-n_p$. We revisit sum rules for vector meson masses at 
finite nucleon density to point out the numerical importance of the screening 
term in the isoscalar channel, which turns out to be one order of 
magnitude larger than any density-dependent condensates over the Borel window. 
This changes the conclusions about
the density dependence of $m_\omega$, indicating $\sim 40$ MeV increase at 
nuclear saturation density. 
\end{abstract} 


\section{Introduction}

Changes of hadronic properties in hot and dense nuclear medium are an 
intriguing issue which ties together modern particle and
nuclear physics. The interest to these questions has been intensified 
over the past decade due to the possibility of studying the transition from 
hadrons to the deconfining phase at heavy ion collisions. 
In particular, the modification of vector meson properties in nuclear 
medium has been a subject of a persistent theoretical 
activity \cite{Review}. This was initiated 
by the idea that in nuclear medium the vector meson masses should drop
as a precursor to the chiral symmetry
restoration \cite{Brown91}.  
Several experiments have also been proposed to study 
the changes of masses, widths and coupling
constants of vector resonances 
in dense (and/or hot) nuclear matter \cite{dileptons}. 

The properties of vector resonances in vacuum and the  
effects of isospin symmetry violation on the 
mixing of the $\rho$, $\omega$  resonances in vacuum have 
been investigated rather carefully by means 
of QCD sum rules in the past \cite{SVZ,SVZII,SVZIII}. 
In the pioneering work of Ref.\,\cite{SVZIII} it was found that 
the nonzero value for the $\rho-\omega$ mixing can be linked to the 
difference of light quark masses, and the possibility of $m_u=0$ is
seemingly excluded.

Later, the QCD sum rule method was extended to finite temperatures 
and densities \cite{Bok}. A number of analyses \cite{HatsLr,HatsLSr,Asak,Jin}
have found that the masses of $\rho$ and $\omega$ resonances decrease in 
nuclear medium\footnote{See, however, the work of Y. Koike \cite{Koike}, where opposite 
behaviour is claimed. A later analysis, based on the 
relation between the current-nucleon forward 
scattering amplitude and the scattering length of the vector 
meson off the nucleon in the static limit, again 
revealed negative mass shifts in the linear density 
approximation \cite{Koike2}.}. In Refs. \cite{Leu,Kli} finite 
widths of the vector mesons have been taken into 
account by hand and by calculation of the $rho$-meson self-energy in 
a chiral model for the 
spectral function, respectively. For the $\rho$-meson channel 
it was found in Ref.\,\cite{Leu} 
that at nuclear saturation density 
an increasing width of the $\rho$-resonance 
necessitates an increasing $\rho$-meson mass. However, for large 
values of the width the mass is blurred 
over a large window of possible values.  
 
While appreciable efforts have been directed to
estimate the density dependent modification of the masses and lifetime 
of the light vector mesons at finite density (and/or temperature),
the question of $\rho-\omega$ mixing at finite 
densities (and/or temperature) has not received much attention. In fact 
finite nuclear densities can have a significant impact on this 
amplitude. The fact that nuclear matter can intrinsically  be
isospin asymmetric implies that the 
$\rho-\omega$ mixing in matter can potentially be larger than 
the vacuum part of the mixing which is induced by the difference in 
$u$ and $d$ quark masses, small in units of characteristic hadronic scales.
This idea was suggested first in Ref.\,\cite{Dutt} 
where it has been pointed out that the presence of 
asymmetric nuclear matter has a profound 
effect on the mixing of the $\rho$ and $\omega$ resonances. 
There the mixing angle was determined from the 
matter induced non-diagonal self energy of the 
$\rho^0$ resonance by employing an SU(2)$_F$ 
symmetric hadronic model. 
Subsequently such a matter induced mixing has also
been analysed on a more elaborate footing in
Ref.\,\cite{broniowski98}.
Along the same lines 
the author of Ref.\,\cite{Bhatt} investigated the nucleonic density and 
temperature dependent $\rho^0$-$\omega$ mixing at a fixed asymmetry. Thereby, 
an enhancement of the modulus of the vacuum mixing amplitude 
was found due to finite density. In all model descriptions the vacuum 
part of the mixing serves as an input parameter, to which all the results are 
normalized by hand in the limit of vanishing density. 
To this end, it is desirable to obtain an independent analysis of the 
mixing using finite density QCD sum rules, which hopefully would allow 
to treat the vacuum part and the density part from the first principles. 
At this point we already 
notice one principal problem of finite density QCD sum rules. 
In the presence of nuclear matter there exist nonscalar condensates
which can be related to the twists of different dimension. In general, 
going from mass dimension $2n$ to 
$2(n+1)$ the ratio of contributions $R^{2n}_{2k}$ with a 
nonzero, fixed twist $2k$ is 
\eqb
R^{2n}_{2k}\propto \frac{A_{2(k+1)}}{A_{2k}}\,\left(\frac{m_N}{M}\right)^2\ .
\eqe
This requires the external momenta 
to be much larger than $m_N$ for the OPE to converge. However, the possibility to 
link properties of a ground state resonance to 
nonperturbative effects in the vacuum (the condensates) via the sum rule 
requires external momenta of $\sim m_N$. We will later show that for the 
contribution of twist operators there is a numerical suppression 
in the corresponding Wilson coeffcients up to mass dimension six. 
Since at higher dimensions we 
have no parametrical smallness the above 
should limit the applicability of QCD sum rules at 
finite nucleonic density.    


In this paper we study the behavior of the isoscalar-isovector 
mixed correlator of the two vector currents in order to extract 
nuclear density effects. The asymptotic behaviour of this correlator 
at large space-like external momenta can be studied within the 
perturbative QCD framework, with the power corrections 
represented by quark, gluon, quark-gluon, etc. condensates.  
In the presence of finite nucleon density the  power correction due to
these condensates will change as compared to their vacuum values. Due to the 
presence of the preferred reference frame, in which the nucleons are at rest, 
new density-dependent power corrections will appear. 
In both cases we assume the small density regime and keep only the 
linear terms in the external nuclear density. As we shall 
see, this approximation is justified for densities not larger 
than the nuclear saturation density, which is small in 
proper ``vacuum'' units. 
    
The asymptotics of the two-point correlation function, calculated this way, 
can be related to the ``phenomenological part'' which
includes the contributions of vector resonances, continuum and the 
screening terms \cite{Bok}. A success or a failure of the QCD sum 
rule analysis of vector meson properties would depend on 
how reliably the contribution of individual resonances 
($\rho$, $\omega$,...) can be separated from the rest of 
the contributions.

We carefully examine the density-dependent part of the 
operator product expansion (OPE) and
find that the effects of matter-induced mixing 
due to nucleonic matrix elements of nonscalar and scalar QCD operators  
in asymmetric nuclear matter follow a certain hierarchy. The asymmetric 
density-induced effects start dominating vacuum contribution
at asymmetries $\alpha_{pn}\equiv (n_p-n_n)/(n_p+n_n)\approx 0.2$ and
an overall nucleonic density twice
the nuclear saturation density $n^0_N=0.17$ fm$^{-3}=$(111\,MeV)$^3$.
However, the analysis of the phenomenological part of the 
QCD sum rules shows that the scattering contribution, 
usually called screening term, turns out to be numerically 
by far more important.  
Brought to the OPE-side of the sum rule, the screening term 
can be regarded as a mass dimension two power correction.  
Already at intermediate asymmetries 
$\alpha_{pn}\approx 0.1$ and saturation density   
vacuum and matter induced $\rho$-$\omega$ mixing 
are of the same sign and comparable in magnitude. 

The smallness of the density-dependent pieces in the OPE as compared to the 
screening term indicates that any conclusion about the density-dependent
piece in the $\rho-\omega$ mixing amplitude will mostly depend on the 
assumptions made about the spectral density, i.e. what is usually called 
phenomenological part of the sum rules. 
This casts strong doubts on the applicability of the finite density 
QCD sum rules for the extraction of the isovector-isoscalar mixing
since the density-dependent ``QCD input'' is negligibly small.
This concern lead us to re-examine the screening terms in the 
isovector-isovector and 
isoscalar-isoscalar correlators which were used in previous works 
\cite{HatsLr,HatsLSr,Asak,Jin} to 
investigate the modification of $\rho$ and $\omega$ masses  
in nuclear matter. We have found that all previous analyses have used
the same value for the screening terms in 
the isovector-isovector and isoscalar-isoscalar correlators. 
This is an unfortunate error because the screening term in the omega
channel turns out to be 9 times larger than the value used in Refs. 
\cite{HatsLr,HatsLSr,Asak,Jin}. This changes
drammatically all the conclusions about the 
behaviour of $m_\omega$ in nuclear matter, and indicates that 
$m_\omega$ is a growing
function of density in the linear density approximation. 


\section{Isosinglet-isotriplet correlator of the two vector currents at 
finite densities}

We start with the (causal) mixed correlator of isotriplet and isosinglet
currents in asymmetric nuclear matter 
\eqb
\label{cor}
\Pi_{\mu\nu}\equiv i\int d^4x\,
\e^{iqx}\la \mbox{T}j_\mu^T(x)j_\mu^S(0)\ra_{n_N}\ ,
\eqe
where 
\eqb
\label{curr}
j_\mu^T\equiv\frac{1}{2}
\left(\bar{u}\gamma_\mu u-\bar{d}\gamma_\mu d\right)\ ,\ \ \ 
j_\mu^S\equiv\frac{1}{2}
\left(\bar{u}\gamma_\mu u+\bar{d}\gamma_\mu d\right)\ .
\eqe
We choose the same normalization of the two currents, which also means that
their couplings to 
physical $\rho$ and $\omega$ resonances are approximately 
equal. In Eq.\,(\ref{cor}) the Gibbs average $\la\ \ra_{n_N}$ 
($n_N$ indicating finite nucleon density) is 
approximated by a vacuum and one-particle nucleon 
states \cite{HatsLr,HatsLSr,Asak,Jin}. 
Due to the presence of a singled out rest frame with four-velocity $u_\mu$
 there are, in general, two independent, current conserving 
tensor structures (longitudinal and isotropic) into which 
$\Pi_{\mu\nu}$ can be decomposed. 
However, in the limit $\vec q\rightarrow 0$ 
one of the corresponding invariants $\Pi_l,\,\Pi_i$ 
becomes redundant \cite{HatsLr}, and we therefore concentrate on 
$\Pi_l$ which satisfies the following 
dispersion relation \cite{HatsLr}
\eqb
\Pi_l(Q_0^2\equiv -q_0^2)\equiv\frac{\Pi^\mu_\mu}{3Q_0^2}=
\frac{1}{\pi}\int_0^\infty ds\, \frac{\mbox{Im}\Pi_l}{s+Q_0^2}+
\mbox{subtractions}\ .
\eqe
Subtracting the terms attributed to the 
$\rho\rightarrow\gamma\rightarrow\omega$ 
electromagnetic mixing from the spectral representation \cite{SVZIII} and 
the OPE and appealing to the literature on density dependent OPE's of 
$\rho$ and $\omega$ current correlators we arrive at the following sum rule 
\eqb
\label{sr}
\Pi^\prime_l(Q_0^2)=\frac{1}{\pi}\int_0^\infty ds\, 
\frac{\mbox{Im}\tilde{\Pi}_l(s)}{s+Q_0^2}+
\mbox{subtractions} ~.
\eqe

The asymptotic behaviour of the lhs of Eq.\,(\ref{sr}) can be calculated 
by means of the operator product expansion (OPE). The result is given in terms 
of the perturbative contribution and power corrections, proportional to the 
condensates, taken in the presence of the external nucleon density. Retaining 
terms up to the order $Q_0^{-6}$, we present the result in the following
form:
\eab
\label{OPE1}
\Pi^\prime_l(Q_0^2)=
-\frac{\alpha}{16\pi^3}\frac{1}{4}\ln Q_0^2+
\frac{1}{Q_0^2}~\frac{3}{2\pi^2}\frac{m_d^2-m_u^2}{4}
+\nonumber\\  \nonumber
\frac{1}{Q_0^4}\left[\frac{m_u}{2}\la\bar{u}{u}\ra_{n_N}-\frac{m_d}{2}
\la\bar{d}{d}\ra_{n_N}
+\frac{2}{3}\fr{Q_\mu Q_\nu}{Q^2} \la{\cal S}(\bar{u}\gamma_{\mu}\cD_{\nu} u
-\bar{d}\gamma_{\mu}\cD_{\nu} d)\ra_{n_N}
\right]-\\ 
\frac{1}{Q_0^6}\left[\frac{\pi\alpha_s}{2}\la
(\bar{u}\gm\gf\lam^a{u})^2-(\bar{d}\gm\gf\lam^a{d})^2
+\fr{2}{9}[(\bar{u}\gm\lam^a{u})^2-(\bar{d}\gm\lam^a{d})^2]\ra_{n_N}
+\right.\\\left.\nonumber
2\pi\alpha\la
\fr{4}{9}(\bar{u}\gm\gf{u})^2-\fr{1}{9}(\bar{d}\gm\gf{d})^2
+\fr{2}{9}[\fr{4}{9}(\bar{u}\gm{u})^2-\fr{1}{9}(\bar{d}\gm{d})^2]\ra_{n_N}+
\right.\\\left.
\frac{8}{3}\fr{Q_\mu Q_\nu Q_\lam Q_\si}{Q^4}
\la{\cal S}(\bar{u}\gamma_{{\mu}}D_{\nu}D_{\lam}D_{\si}u
-\bar{d}\gamma_{{\mu}}D_{\nu}D_{\lam}D_{\si}d)\ra_{n_N}\right]\ .\nonumber
\eae
In Eq.\,(\ref{OPE1}), symbol ${\cal S}$ denotes the operation of
making tensors symmetric and traceless. As usual, 
(for example \cite{HatsLr,Leu,Koike,Kli}), the averages over mixed operators 
and twist four contributions have been omitted in Eq.\,(\ref{OPE1}). 
The former can either be reduced to four quark operators 
by use of the equation of motion 
(these contributions are already included), or they are suppressed at 
$\mu^2\approx 1$ GeV$^2$ since 
there the gluon content of the nucleonic wave function 
is small \cite{HatsLr}. The latter has been argued 
in Ref.\,\cite{HatsLSr} to have no 
substantial effect on 
the $\rho$ and $\omega$ mass shifts, 
and we will therefore omit twist four operators. 
Further progress in calculating
the OPE depends on how accurately we can predict the size of 
various contributions to $\Pi'$. We restrict ourselves to the case of
low and medium densities, so that the linear (mean field) 
approximation is justified, and the density dependent part enters in the 
final expression multiplied by the matrix elements over the 
single nucleon states. We further make use of the vacuum saturation hypothesis 
\cite{SVZ}, which becomes an exact relation in the limit of large number
of colors. This hypothesis is known to ``work'' reasonably well in 
vacuum. However, its application to the nucleon matrix elements is not
fully justified. We use this hypothesis to estimate the order of
magnitude of dim 6 contribution, noting that their numerical weight in 
the final result turns out to be small as the OPE is largely dominated 
by dim 4 contributions. With all these assumptions, Eq. (\ref{OPE1}) 
can be reduced to the following form:
\eab
\label{OPE'}
\Pi^\prime_l(Q_0^2)=
-\frac{\alpha}{16\pi^3}\frac{1}{12}\ln Q_0^2+
\frac{1}{Q_0^2}\left[\frac{3}{2\pi^2}\frac{m_d^2-m_u^2}{12}-
\right]+\nonumber\\  
\frac{1}{Q_0^4}\left[\frac{m_u-m_d}{2}\{\la\bar{q}{q}\ra_0(\mu^2)+
\frac{\Sigma_{\pi N}(\mu^2)}{2\bar{m}}\,n_N\}-
\bar{m}\la p|\bar{u}u-\bar{d}d|p\ra_{\vec{k}_p=0}\alpha_{pn}n_N+
\fr{1}{2}m_p \alpha_{pn} n_N A_2^{u-d}(\mu^2)\right]+\nonumber\\ 
\frac{1}{Q_0^6}\left[-\frac{112}{81}\pi[\alpha_s\la\bar{q}{q}
\ra_0^2(\mu^2)\{-\gamma+\frac{\alpha}{8\alpha_s(\mu^2)}\}+
2\la\bar{q}{q}\ra_0\la p|\bar{u}u-\bar{d}d|p\ra_{\vec{k}_p=0}\alpha_{pn}n_N]-
\frac{5}{12}m_p^3\alpha_{pn} n_N A_4^{u-d}(\mu^2)\right]\ .
\eae
In Eq.\,(\ref{OPE'})
$n_N$ ($n_p$, $n_n$) denotes the total 
nucleonic (proton, neutron) density; $\alpha_{pn}$ the $p$--$n$ asymmetry, 
defined as $\alpha_{pn}\equiv (n_p-n_n)/n_N$; $m_p$ the proton mass; 
$\la\bar{q}{q}\ra_0$ the value of the $u$-quark condensate; 
$\gamma$ the asymmetry of $u$-quark and $d$-quark condensate defined as
$\gamma\equiv\la\bar{d}{d}\ra_0/\la\bar{u}{u}\ra_0-1$; 
$\Sigma_{\pi N}=(45\pm 7)$ MeV 
\cite{HatsLSr} the nucleon 
sigma term; and $\bar{m}$ is defined as 
$\bar{m}\equiv 1/2(m_u+m_d)$. The electromagnetic 
coupling is $\alpha$, the strong coupling at scale 
$\mu$ is $\alpha_s(\mu^2)$. In Eq. (\ref{OPE'}) we have already used
the numerical smallness of isospin and chiral 
symmetry violating parameters as compared
to the normal hadronic scale and thus neglected terms proportional to
$m_{u(d)} \gamma$, $\gamma^2$ and so on. For similiar reasons it is 
justified to neglect the effects of isopsin violation in nucleon matrix 
elements and take $\la p|\bar{u}u-\bar{d}d|p\ra = - 
\la n|\bar{u}u-\bar{d}d|n\ra$, which leads to the dependence on the 
asymmetry factor $\alpha_{pn}$. The scalar matrix element 
$\la p|\bar{u}u-\bar{d}d|p\ra$ is related to the baryon 
octet mass splitting, $(m_\Xi-m_\Sigma)/m_s$, and it is numerically close to 
0.7 \cite{Zh}. 

As for the 
contribution of symmetric and traceless 
twist two quark bilinears \cite{HatsLSr} of dimension four and six, 
their nucleonic matrix elements are determined 
by the quark parton distributions $A_2^{u-d}(\mu^2)$ and 
$A_4^{u-d}(\mu^2)$ as \cite{HatsLr}
\eab
\label{tw2}
\la{\cal S}\bar{q}\gamma_{\mu_1} 
D_{\mu_2} q\ra_{N(k)}(\mu^2)&=&
-i A_2^{u-d}(\mu^2)\left(k_{\mu_1} k_{\mu_2}-
\frac{1}{4}g_{{\mu_1}{\mu_2}}k^2\right)\ ,\nonumber\\ 
\la{\cal S}\bar{q}\gamma_{{\mu_1}} D_{\mu_2}
D_{\mu_3}D_{\mu_4}q\ra_{N(k)}(\mu^2)&=&
i A_4^{u-d}(\mu^2)\left(k_{\mu_1} k_{\mu_2} k_{\mu_3} 
k_{\mu_4}-\mbox{traces}\right)\ ,
\eae
where $k_\mu$ denotes the nucleon momentum, 
$D_\mu$ is the gauge covariant derivative. In general, the factor 
$A_k^{q}(\mu^2)$ can be obtained from the parton
distributions $Q(x,\mu^2)$ and $\bar{Q}(x,\mu^2)$ in the proton as 
\eqb
\label{A}
A_k^{q}(\mu^2)=2\int_0^1 dx\, x^{k-1}\left(Q(x,\mu^2)+
(-)^k\bar{Q}(x,\mu^2)\right)\ . 
\eqe
In Ref.\,\cite{Glueck} the parton distributions in the 
nucleon have been fitted to experiment 
at a resolution scale $\mu^2=0.26$ GeV$^2$. 
Using these distributions and performing the integrations 
of Eq.\,(\ref{A}), we obtain
\eqb
\label{Ac}
A_2^{u-d}(\mu^2=0.26\ ,\mbox{GeV}^2)=0.429\ ,\ \ \ 
A_4^{u-d}(\mu^2=0.26\ ,\mbox{GeV}^2)=0.097\ .
\eqe
To generate the respective values at 
the scale $\mu^2\approx 1$ GeV$^2$ relevant for the 
sum rule we simply use the conversion factors $f_2(\mu_2^2,\mu_1^2)\equiv
\frac{A_2^{u+d}(\mu_2^2)}{A_2^{u+d}(\mu_1^2)}$ and $f_4(\mu_2^2,\mu_1^2)\equiv
\frac{A_4^{u+d}(\mu_2^2)}{A_4^{u+d}(\mu_1^2)}$. 
Using $A_2^{u+d}$ and $A_4^{u+d}$ at 1 GeV from Ref.\,\cite{HatsLSr}, 
we arrive at the following values of matrix elements of interest
\eqb
A_2^{u-d}(1\,\mbox{GeV}^2)=0.32\ ,\ \ \ 
A_4^{u-d}(1\,\mbox{GeV}^2) = 0.062\ .
\eqe
Keeping this in mind, we will for now proceed to the numerical evaluation
of OPE. Performing a Borel transformation of the OPE of Eq.\, (\ref{OPE'}) and 
omitting the numerically strongly suppressed dimension two power correction, 
we obtain
\eab
\label{OPE'M}
\Pi^\prime_l(M^2)=
3.7\times 10^{-6}+
\fr{|\qq|}{M^4}\left(2\,{\rm MeV} - 1.5 \,{\rm MeV}~
\fr{\alpha_{np}}{0.2}~\frac{n_N}{n^0_N}\right)-\\\nonumber
\fr{|\qq|}{M^4}\fr{0.1 \,{\rm GeV}^2}{M^2}
\left(1.4\,{\rm MeV} + \fr{\alpha_{np}}{0.2}~\frac{n_N}{n^0_N}~
[3.8 \,{\rm MeV}- 2.4 \,{\rm MeV}]
\right)
\eae
In this expression we have used the following set of values:\\  
$m_u=5\,$MeV; $m_d=9\,$MeV; $\bar{m}=7\,$MeV; $m_p=940\,$MeV; 
$n_N=n^0_N=(111\,$MeV)$^3$ (the nuclear matter saturation density); 
$\Sigma_{\pi N}=45\,$MeV; $\la\bar{q}{q}\ra_0=-(225\,$MeV)$^3$; 
$\gamma=-10^{-2}$ \cite{HatsHMK}; $\alpha=1/137$; and 
$\alpha_s(1\,\mbox{GeV}^2)=0.5$ \cite{HatsLSr}. 
The quark condensate has been factored out numerically 
for the sigma-term and the twist contributions. 

Several important observations should be made at this point. For 
$M \sim 1 $ GeV, a pure perturbative contribution is negligibly small 
as compared to power corrections. The latter are dominated by 
dimension 4, with the constant term originating from $m_u$ and $m_d$ mass 
difference and the $\alpha_{np}n_N$-dependent piece given by the $A_2^{u-d}$
contribution. At the level of dimension 6 we observe three different 
terms (second line of Eq. (\ref{OPE'M})): vacuum part, 
density-dependent scalar condensate and 
twist contributions. At this dimension the density dependent 
contribution from scalar condensate and twist tend to cancel
each other. This cancellation can be an artefact
of chosen parameters and/or of the crude nature of approximations made in
estimating the size of the four-quark matrix elements over the nucleon.
Nevertheless, $M \sim 1 $ GeV, and the OPE is dominated by dim 4 terms, where 
at $\fr{\alpha_{np}}{0.2}~\frac{n_N}{n^0_N} \sim 1$ the suppression of 
dim 6 is about 50\%. For higher values of asymmetric density
$\Pi'(M^2=1\,{\rm GeV})$ changes sign. 

What does this behaviour of $\Pi'(M^2=1\,{\rm GeV})$ mean in terms of 
the $\omega-\rho$ resonance mixing amplitude?
To answer this question we should parametrize the spectral function 
in terms of the resonance contributions and analize 
the resulting sum rule (\ref{sr}). 

Following Refs. \cite{SVZIII,Bok}, we approximate the imaginary part of the 
correlator by contributions of $\rho$, $\omega$, $\rho'$, $\omega'$ resonances
and continuum:
\eqb
\label{Pit}
\frac{1}{\pi}\mbox{Im}\tilde{\Pi}_l(s,\alpha_{pn},n_N)=\frac{1}{4}
\left[f_\rho\delta(s-m_\rho^2)-f_\omega\delta(s-m_\omega^2)+
f_{\rho^\prime}\delta(s-m_{\rho^\prime}^2)-f_{\omega^\prime}
\delta(s-m_{\omega^\prime}^2)
+ \fr{\rho_{sc}^{ST}}{8\pi^2}\delta(s)+
\frac{\alpha}{16\pi^3}\theta(s-s_0)\right].
\eqe 
The contribution to the mixing due 
to the electromagnetic continuum is small \cite{SVZ}. 
Therefore, we will neglect it in the subsequent 
consideration. 
In Eq.\,(\ref{Pit}) $f_\rho$  and $f_\omega$ refer to the $\rho$  and 
$\omega$ residues of the $\rho$-$\omega$ current propagator, 
and $\rho^\prime$  and $\omega^\prime$ symbolize the cumulative 
effect of higher resonances\footnote{In Ref.\,\cite{SVZ} the cumulative values 
$m^2_{\rho^\prime},m^2_{\omega^\prime}$ were chosen to be about 1.5 GeV$^2$, 
which is well below the physical masses 
($\sim$ 1.7 GeV) of the resonances $\rho^\prime,\omega^\prime$.} 
introduced in the original analysis
\cite{SVZIII} in order to have consistent asymptotic behaviour of $\Pi'(M^2)$. 
Besides the ``usual'' annihilation continuum above a certain threshold $s_0$,
Eq. (\ref{Pit}) exhibits a scattering term which behaves as a pole at 
$s=0$ (Landau pole) \cite{Bok}. The corresponding coefficient can 
be calculated explicitly, and the result in the leading order in 
Fermi momentum ($p_f/m_N$ expansion) is given by
\eqb
\rho_{sc}^{ST} = \fr{2\pi^2}{m_N}\left[ F^S_pF^T_p n_p + F^S_nF^T_n n_n
\right]= \fr{6\pi^2}{m_N}\alpha_{pn}n_N.
\label{scrST}
\eqe
Here the coefficients $F_{p(n)}^{S(T)}$ are defined via nucleon matrix 
elements of quark vector currents at vanishing momentum transfer:
\eab
F_p^S=F_n^S=\la p|\bar u \gamma_0 u + \bar d \gamma_0 d|p \ra = 3\nonumber\\
F_p^T=-F_n^T=\la p|\bar u \gamma_0 u - \bar d \gamma_0 d|p \ra = 1.
\eae
After Borel transformation the contribution of the Landau 
screening term is usually carried to the lhs of the sum rule 
to effectively become a power correction of 
dimension two in the expansion of $\Pi^\prime_l(M^2)$. Defining 
$f_{\rho\omega}\equiv 1/2\,(f_\rho+f_\omega)$, 
$\bar{m}_r^2\equiv 1/2\,(m_\rho^2+m_\omega^2)$, 
$\Delta m_r^2\equiv m_\omega^2-m_\rho^2$, and 
$\beta=(f_\omega-f_\rho)\bar{m}_r^2/(f_{\rho\omega}\Delta m_r^2)$ 
(the primed quantities are defined analogously), we quote 
the result of Ref.\,\cite{SVZIII} relating $f_{\rho\omega}$ to the 
measurable quantities $\bar{m}_r^2$, $\Delta m_r^2$, $g_\rho$, and $g_\omega$
\eqb
f_{\rho\omega}\approx-\frac{12\bar{m}_r^2}{g_\rho g_\omega}
\frac{\delta_{\rho\omega}}{\Delta m_r^2}
\equiv \frac{m_r^4}{\Delta m_r^2}\xi\ ,
\eqe
where $g_\rho$, $g_\omega$ are the respective decay constants, 
and $\delta_{\rho\omega}$ enters the measurable 
mixing parameter $\ep$ as follows
\eqb
\label{mixep}
\ep=\frac{\delta_{\rho\omega}}{(m_\omega-1/2i\Gamma_{\omega})^2-
(m_\rho-1/2i\Gamma_\rho)^2}\ .
\eqe
Thereby, $\ep$ is defined as
\eqb
\omega=\omega_0+\ep\rho_0\ ,\ \ \ \ \ \rho=\rho_0-\ep\omega_0\ , 
\eqe
and $\Gamma$ denotes the width of the respective resonance. It is fair to 
remark at this point that the observable combination, 
$\ep\simeq\delta_{\rho\omega}\Gamma_\rho^{-1}m_\rho^{-1}$ will have 
an additional dependence on density due to a substantial increase of
$\Gamma_\rho$ with $n_N$ \cite{gamma}. Thus, finding the decrease of
$\xi$ with density would certainly allow to conclude that $\ep$ is
decreasing. The opposite behavior, a rising $\xi$, 
would complicate the prediction 
of $\ep(n_N)$.

The final sum rule is given by the following expression:
\eab
\label{final}
\nonumber \fr{1}{4}\xi\frac{\bar{m}_r^2}{M^2}
\left(\frac{\bar{m}_r^2}{M^2}-\beta\right)\,\e^{-\bar{m}_r^2/M^2}+
(\rho\rightarrow\rho^\prime,\, \omega\rightarrow\omega^\prime)=\\
1.1\cdot 10^{-2} \,{\rm GeV}^{-1}\left\{
\fr{18 \,{\rm MeV}}{M^2}\fr{\alpha_{np}}{0.2}~\frac{n_N}{n^0_N}
+\fr{1}{M^4}\left(2\,{\rm MeV}- 1.5 \,{\rm MeV}~
\fr{\alpha_{np}}{0.2}~\frac{n_N}{n^0_N}\right)- \right.\\\nonumber\left.
\fr{0.1}{M^6}
\left(1.4\,{\rm MeV} + \fr{\alpha_{np}}{0.2}~\frac{n_N}{n^0_N}~
[3.8 \,{\rm MeV}- 2.4 \,{\rm MeV}]\right)\right\}
\nonumber
 ,
\eae
where all masses and the Borel parameter are taken in units of GeV. 
It is remarkable that the screening term, brought to the OPE side of 
this sum rules, completely dominates other density dependent contributions.
This shows that in the asymmetric 
nuclear matter background the influence of the 
screening term on the $\omega-\rho$ mixing is by far 
more important than any changes of the QCD condensates. 
Moreover, for any realistic $M^2$ the screening term becomes comparable to 
the vacuum contribution to the 
mixing at $n_N\simeq n_N^0$ and asymmetries as low as 
$\alpha_{np}\sim 0.05$. 

In the limit of vanishing density, relation (\ref{final}) reduces to the 
known sum rule for $\rho-\omega$ mixing. A naive evaluation of this sum rule 
at $\rho$-meson mass, $M^2=(0.77)^2$, and at $n_N=0$, gives a reasonable
agreement with experimentally measured value $\xi= 1.1\times 10^{-3}$ 
\cite{Barkov} with $\beta\simeq 0.5$, advocated in Ref. \cite{SVZIII}. 
Next, we parametrize the linear dependence of $\xi$ and $\beta$ on the density 
as follows:
\eab
\xi = \xi^{(0)} +  \xi^{(1)}\tilde n;\;\;\;\;\;\;
\beta = \beta^{(0)} +  \beta^{(1)}\tilde n,
\eae
where $\tilde n$ denotes $\alpha_{np}n_N$ in units of $0.2n_N^0$.

The primary reason for the introduction of the $\rho'-\omega'$ contribution
in Ref. \cite{SVZIII} was the absence of the $1/M^2$ term in the 
OPE side of the sum rule, so that $\rho$ and $\omega$ contribution alone
would not be consistent with the asymptotic behaviour of $\Pi'$. Thus, 
the role 
of  $\rho'-\omega'$ is to imitate the cancellation of $1/M^2$ terms in 
contributions of various resonances 
at large $M^2$. For a semiquantitative determination of the linear density 
dependence of 
$\xi$ and $\beta$ we proceed as in Ref.\,\cite{SVZ}. 
There the vacuum values of 
$\xi$ and $\beta$ were estimated by choosing $M=m_\rho$, which strongly 
suppresses 
the higher resonances. It should then be legitimate to compare powers of 
$M^{-2}$ 
in the OPE and the lowest 
resonance contribution. The result is given by  
\be
\fr{\xi^{(1)}}{\xi^{(0)}}+\fr{\beta^{(1)}}{\beta^{(0)}}=-2.0\cdot 10^{-4}~
\fr{4}{\xi^{(0)}\beta^{(0)}\bar m_r^2}\ .
\ee
Using this relation, we can find $\beta^{(1)}$ and $\xi^{(1)}$ separately,
evaluating (\ref{final}) at $M^2=0.59$. The final estimate of $\xi^{(1)}$
reads as
\eqb
\xi^{(1)} \simeq [2.3 - 0.8]\times 10^{-3}= 1.5 \times 10^{-3},
\eqe
where 2.3 originates from the screening term and -0.8 comes from the OPE. 
A similar number can be obtained from the combination of (\ref{final}) and its
first derivative in $M^2$. This value of $\xi^{(1)}$ leads to the doubling 
of mixing amplitude and complete screening 
at $n_n-n_p \sim \pm 2.5 \times 10^{-2} ~{\rm fm^3}$, respectively.

\section{Importance of the screening term for the isoscalar-isoscalar 
correlator} 
 
Having found such an important role of the screening term 
in the isoscalar-isovector mixed correlator, we would like to return 
to previous analyses of diagonal correlators (isovector-isovector and
isoscalar isoscalar) which were used to extract the behaviour of $m_\rho$ and
$m_\omega$ at finite nucleon density \cite{HatsLr,HatsLSr,Asak,Jin}. 
In all these papers it was found that masses and coupling constants
of $\rho$ and $\omega$ resonances behave similarly in nuclear matter, simply
because the OPE sides of the sum rules in both cases are the same 
after the application of the vacuum saturation hypotheses. 

We use the same symmetric normalization of the two currents, 
Eq. (\ref{curr}). From now on we neglect the asymmetry of the 
nuclear matter and other isospin breaking effects. Then 
the sum rules for isovector-isovector and isoscalar-isoscalar correlators
in medium take the following symbolic form:
\eab
\label{rho}
\fr{1}{M^2}F_\rho^*e^{-m^{*2}_\rho/M^2} = 
\fr{1}{8\pi^2}\left(1-e^{-S_\rho^*/M^2}\right)
-\fr{1}{4}\fr{n_N}{m_NM^2} + \fr{c_4}{M^4} + \fr{c_6}{2M^6} \\
\fr{1}{M^2}F_\omega^*e^{-m^{*2}_\omega/M^2} = 
\fr{1}{8\pi^2}\left(1-e^{-S_\omega^*/M^2}\right)
-\fr{9}{4}\fr{n_N}{m_NM^2} + \fr{c_4}{M^4} + \fr{c_6}{2M^6},
\label{omega}
\eae
where $c_4$ and $c_6$ are the same for both expressions. 
Obviously, at vanishing nucleon density $F_\omega \simeq F_\rho$ and
$S_\omega\simeq S_\rho$. 

It is remarkable, that the screening terms in Eqs. (\ref{rho}-\ref{omega}) are 
different by a factor of 9. The enhancement of the screening term 
in the  isoscalar-isoscalar channel is due to 
\eab
\fr{\rho_{Sc}^{SS}}{\rho^{TT}_{Sc}}=\fr{F^S_pF^S_p}{F^T_pF^T_p} = 9 \ .
\eae
This difference was overlooked in Refs. \cite{HatsLr,HatsLSr,Jin}
\footnote{In Ref. \cite{Jin} $\rho_{sc}$ was taken as a free search parameter 
and determined {\em from} the sum rules at the level consistent with 
$\rho^{TT}_{Sc}$ for {\em both} correlators. 
It casts a strong doubt on the validity of the whole approach,
since the actual value of the screening term for $\omega$ should be 9 times 
larger.}.

The coefficients $c_2$ and $c_4$ can be computed along the same standard
technique (again with considerable degree of uncertainty for 
$c_6$). When plugging these values into the sum rules 
(\ref{rho}-\ref{omega}), we obtain the following
numerical relations:
\eab
\label{rho1}
\fr{1}{M^2}F_\rho^*e^{-m^{*2}_\rho/M^2} = 
\fr{1}{8\pi^2}\left(1-e^{-S_\rho^*/M^2}\right) - \fr{3.4\times 10^{-4}}{M^2}~
\fr{n_N}{n_N^0}+
\fr{10^{-4}}{M^4}\left[4.1+3.8~\fr{n_N}{n_N^0}\right]
 + \fr{10^{-4}}{2M^6} \left[-2.8+1.2~\fr{n_N}{n_N^0}\right]
\\
\fr{1}{M^2}F_\omega^*e^{-m^{*2}_\omega/M^2} = 
\fr{1}{8\pi^2}\left(1-e^{-S_\omega^*/M^2}\right)- 
\fr{31\times 10^{-4}}{M^2}~\fr{n_N}{n_N^0}+
\fr{10^{-4}}{M^4}\left[4.1+3.8~\fr{n_N}{n_N^0}\right]
 + \fr{10^{-4}}{2M^6} \left[-2.8+1.2~\fr{n_N}{n_N^0}\right]
\label{omega1}
\eae
where again all masses and dimensional coupling constants are 
taken in GeV units. It is remarkable that at $M\sim 1 GeV$ 
the screening term in the $\omega$ sum rule is 
larger by an order of magnitude than any other density-dependent term from the OPE!

As in the previous case, it is convenient to 
parameterize the density dependence of 
masses and coupling constants as follows:
\eab
m = m^{(0)}\left(1 +  \fr{m^{(1)}}{m^{(0)}}~\fr{n_N}{n_N^0}\right);
\;\;\;\;\;\;
F = F^{(0)}\left(1 +  \fr{F^{(1)}}{F^{(0)}}~\fr{n_N}{n_N^0}\right);
\;\;\;\;\;\;
S_0 = S^{(0)}_0\left(1 +  \fr{S^{(1)}_0}{S^{(0)}_0}~\fr{n_N}{n_N^0}\right).
\eae
 
Using the sum rules (\ref{rho1}) and (\ref{omega1}), and the first 
derivatives of these expressions, we solve for $m^{*2}$ as a function 
of $S_0^*$, and Borel parameter $M$. The dependence of the threshold on the 
the density is obtained by requiring the Borel curves, $m^*(M^2,S^*,n)$, be parallel 
over the Borel window which we take from 0.6 to 1.2 GeV for different values of densities. 
The slope of the Borel curve $m(M^2)$ in the Borel window at zero density 
represents a ``systematic 
uncertainty'' introduced by sum rules and the requirement of the Borel curves to be 
parallel at different densities is equivalent to the requirement 
that this uncertainty does not change
while going to finite but small densities.  The resulting dependence of 
$S_0$ on the density, $S^{(1)}_0/S^{(0)}_0 = -0.2$ for $\rho$ and $-0.1$ for $\omega$, 
allows us to deduce the following estimates for the linear dependence of masses on 
the density:
\eab
\fr{m_\rho^{(1)}}{m_\rho^{(0)}} \sim -0.15\ , \;\;\;\;\;\;
\fr{m_\omega^{(1)}}{m_\omega^{(0)}} \sim 0.05.
\label{slope}
\eae

Our estimate for $m_\rho^{(1)}$ agrees with the results of previous 
analyses \cite{HatsLr,HatsLSr,Jin}. The result for  $m_\omega^{(1)}$
has the opposite sign and correspond to an $40$ MeV increase of
$m_\omega$ at the nuclear saturation density. This difference could be easily explained 
by the error in the screening term for $\omega$ sum rule in \cite{HatsLr,HatsLSr,Jin}.
The disagreement with the results of \cite{Kli}, where the correct form of the screening 
terms is used, is harder to explain, and we hypothesize that it could be an artefact of 
different numerical methods used to extract the 
density dependence of the resonance masses and thresholds. 
 
\section{Discussion}

Apart from the question of (non)convergence of the OPE 
we would like to point out some concerns about usefulness and 
validity of the sum rules at finite densities. 
\begin{enumerate}

\item{\em  Vacuum factorization at dim 6}. It is unclear what the
status of factorization procedure is, especially in the presence of 
nuclear matter. In principle, one could try to relate four-fermion
matrix elements over the nucleon states, which appear in the 
calculation of the OPE, to some measured processes induced 
by weak interactions.  Indeed, non-leptonic hyperon decays
and parity violating pion-nucleon coupling constants could be reduced 
to similar matrix elements from the four-quark operators. It is 
unclear, though, whether such an analysis is feasible. 

\item {\em The importance of a particular choice of the spectral function}.
In linear density approximation 
the analysis of the examples of the 
$\rho-\omega$ and $\omega-\omega$ sum rules suggest that there are large contributions from 
the respective screening terms. In fact, these contributions dominate 
all density dependent pieces in the OPE. It means that the 
``QCD input'' in these channels is not important in comparison with 
the choice of the spectral function at finite density. 

\item {\em Is the linear density approximation valid up to $n_0$ and beyond?}
The use of the dilute Fermi gas to model 
the behaviour of the scattering terms and the QCD condensates 
has its limitations, and a more realistic description may 
greatly affect the resulting sum rule. However, it seems 
unfeasible to calculate QCD operator averages over 
interacting multi nucleon states which one would have to consider when going 
beyond the dilute gas approximation. An inclusion of Fermi 
motion of noninteracting nucleons is practically doable 
but does not alter the zero momentum result significantly.      

\end{enumerate}

In conclusion, we have considered the correlator of isovector and isosinglet 
vector currents in the presence of the asymmetric nuclear matter in linear density 
approximation. 
We see a
significant dependence of the OPE on $n_n-n_p$, which becomes comparable to 
vacuum contributions at $n_n-n_p \sim 0.05 n^0_N$. An attempt to extract
the $\rho$-$\omega$ mixing, using the dispersion relation has shown that 
this mixing is more affected by the presence of the scattering term 
than by density-dependent part of the OPE. A similar tendency exists 
in the isosinglet-isosinglet channel, which is normally used to deduce the 
dependence of $m_\omega$ on density. Hence, in linear density approximation the 
explicitely density dependent part of the {\em spectral functions} 
(scattering terms) in the $\rho$-$\omega$ and $\omega$-$\omega$ channels dominantly drive 
the density dependence of hadronic parameters. The density dependence of the width 
of the resonances, which has been neglected here, does not alter this finding. However, 
it may drastically change the conclusions about the direction of resonance 
mixing at finite, asymmetric density.

\section*{Acknowledgments}

R.H. and M.P. would like to thank V. Eletsky and A. Vainshtein for sharing
their scepticism about the relevance of QCD sum rules at finite density.
M.P. is grateful to the McGill university nuclear theory group,
where this work was started, for the hospitality 
extended to him during his visit. 
The work of R.H. was funded by a fellowship of Deutscher 
Akademischer Austauschdienst (DAAD). The work of A.K.D.M. has been supported by
the Natural Sciences and Engineering Research Council of Canada and the Fonds
FCAR of the Queb\'ec Government.  The work of M.P. was supported in part by
the Department of Energy under Grant No. DE-FG02-94ER40823.

\bibliographystyle{prsty}

\end{document}